\pdfoutput=1

\documentclass[11pt]{article}

\usepackage{acl}

\usepackage{times}
\usepackage{latexsym}
\usepackage{graphicx}
\usepackage{xurl}
\usepackage{amsmath}
\usepackage{cases}
\usepackage{amssymb}
\usepackage{amsthm}
\usepackage{breqn}
\usepackage{algorithm}
\usepackage{algpseudocode}
\usepackage{enumitem}
\usepackage{caption}
\usepackage{subcaption}

\theoremstyle{definition}
\newtheorem{definition}{Definition}[section]
\newtheorem{theorem}{Theorem}[section]

\newtheorem{lemma}{Lemma}[section]

\DeclareMathOperator*{\argmax}{arg\,max}
\DeclareMathOperator*{\argmin}{arg\,min}

\usepackage[T1]{fontenc}

\usepackage[utf8]{inputenc}

\usepackage{microtype}

\usepackage{inconsolata}

\title{Differentially Private Next-Token Prediction of Large Language Models}

 \author{James Flemings \quad Meisam Razaviyayn \quad Murali Annavaram \\
         University of Southern California \\
         \texttt{\{jamesf17, razaviya, annavara\}@usc.edu }}

\begin{document}
\maketitle

\begin{abstract}
Ensuring the privacy of Large Language Models (LLMs) is becoming increasingly important. The most widely adopted technique to accomplish this is DP-SGD, which trains a model to guarantee Differential Privacy (DP). However, DP-SGD overestimates an adversary's capabilities in having white box access to the model and, as a result, causes longer training times and larger memory usage than SGD. On the other hand, commercial LLM deployments are predominantly cloud-based; hence, adversarial access to LLMs is black-box. Motivated by these observations, we present Private Mixing of Ensemble Distributions (PMixED): a private prediction protocol for next-token prediction that utilizes the inherent stochasticity of next-token sampling and a public model to achieve Differential Privacy. We formalize this by introducing RD-mollifers which project each of the model's output distribution from an ensemble of fine-tuned LLMs onto a set around a public LLM's output distribution, then average the projected distributions and sample from it. Unlike DP-SGD which needs to consider the model architecture during training, PMixED is model agnostic, which makes PMixED a very appealing solution for current deployments. Our results show that PMixED achieves a stronger privacy guarantee than sample-level privacy and outperforms DP-SGD for privacy $\epsilon = 8$ on large-scale datasets. Thus, PMixED offers a practical alternative to DP training methods for achieving strong generative utility without compromising privacy.
\end{abstract}
\section{Introduction}
Large language models (LLMs) are being deployed to improve societal productivity, from troubleshooting complex systems to autocompletion tools and interactive chatbots. Their commercial success is largely attributed to their ability to generate human-like text. However, when given query access to an LLM, it has been shown that LLMs are susceptible to training data extraction attacks \cite{carlini2021extracting} due to memorization of training samples \cite{carlini2019secret}. These security vulnerabilities have recently catalyzed government intervention, most notably the EU's AI Act \cite{EUAIAct2024} and the US executive order on Safe, Secure, and Trustworthy AI \cite{USAct2023}. Thus, it is becoming a requirement that entities deploying LLMs must do so in a privacy-preserving way.

The gold standard for achieving strong privacy is Differential Privacy (DP), a mathematical framework that reduces how much an LLM memorizes individual data samples \cite{dwork2006differential}. The most widely known technique injects strategic noise into the training algorithm, called DP-SGD \cite{abadi2016deep}. It has recently been shown that applying DP-SGD during fine-tuning on pre-trained LLMs provides acceptable results \cite{li2021large, yu2021differentially}. Unfortunately, scaling these results to larger datasets and models becomes challenging since (1) The magnitude of the noise added by DP-SGD scales with the total number of parameters of the model, i.e., $\sqrt{d}$ \cite{kamath2020privml}, in the worst case; and (2) ML accelerated hardware is designed to exploit batch operations but is underutilized by DP-SGD since it requires per-sample gradient calculations, which cause large runtime and memory consumption \cite{yousefpour2021opacus}.

\begin{figure*}[h!]
    \centering
    \includegraphics[width=2.00\columnwidth]{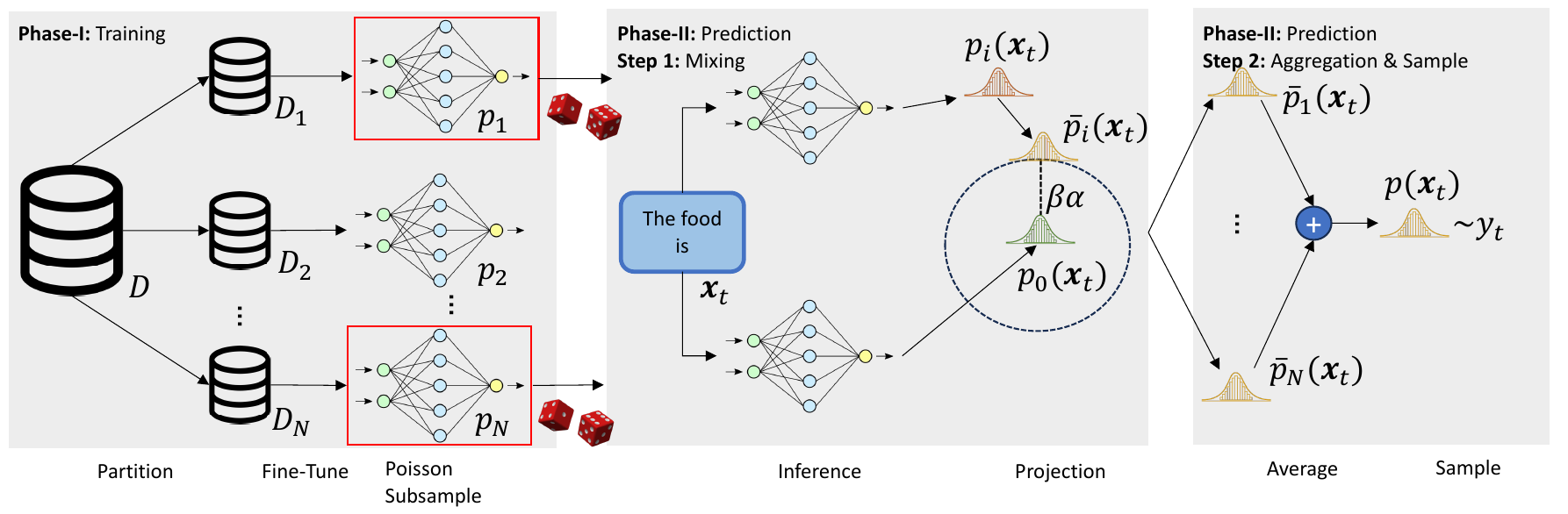}
    \caption{A brief overview of PMixED, which can be broken down into two phases. In Phase-I, the private dataset $D$ is partitioned into $N$ pairwise disjoint subsets $D_1, ..., D_N$, each of which $D_i$ is fine-tuned with a pre-trained LLM to produce $p_{i}$. Afterward, PMixED performs private predictions in Phase-II which can be further broken down into two steps. In Step 1, which we call mixing, a query $\mathbf{x}_t$ from a user is received at time $1 \leq t \leq T$. First, PMixED subsamples a subset of the ensemble, then generates the output distribution of each selected model $p_{i}(\mathbf{x}_t)$ and the output distribution of a public model $p_0(\mathbf{x}_t)$. Each $p_i(\mathbf{x}_t)$ is projected along a Renyi Divergence ball centered at the output distribution $p_{0}(\mathbf{x}_t)$ with radius $\beta\alpha$ to produce $\overline{p}_{i}(\mathbf{x}_t)$, which is a mixture of private $p_i(\mathbf{x}_t)$ and public $p_0(\mathbf{x}_t)$ information. In Step 2, all projected distributions are averaged into $p(\mathbf{x}_t)$ then sampled $y_t \sim p(\mathbf{x}_t)$.}
    \label{fig:PMixED_overview}
\end{figure*}

A key observation that motivates this work is that many commercial deployments of LLMs are only accessible via an API, e.g. Chat-GPT. Hence, an adversary has only black-box access to the model, yet DP-SGD assumes the adversary has white-box access to the model since it applies differential privacy to the model parameters. Such a pessimistic assumption on adversarial capabilities can lead to overestimating the privacy loss bounds \cite{nasr2021adversary}. We believe that improving the privacy of LLMs under the black box assumption is the key to enabling wider adoption of privacy-preserving LLMs. However, prior work has demonstrated that private prediction algorithms generally perform worse than private training algorithms \cite{van2020trade}.

To address the aforementioned limitations, we propose PMixED, depicted in Figure \ref{fig:PMixED_overview}. Rather than providing DP during training, PMixED instead provides DP for a private corpus during inference by utilizing two crucial features: (1) Randomness comes for free when predicting the next-token by sampling from the output probability distribution of a language model; (2) We can bound the privacy leakage of a prediction using a public model\footnote{While it is assumed that a public LLM does not compromise the privacy of a private corpus, in practice, such models can inadvertently leak information. This assumption is implicitly relied upon in works that employ differentially private fine-tuning of public models.} to mix the predictions of privacy-sensitive LLMs. We formalize these two observations by introducing $(\alpha, \beta)$-RD Mollifiers which generalize $\epsilon$-Mollifers introduced in \citet{husain2020local} and could be of interest independent of this work. 

PMixED can be broken down into two phases: training and private prediction. During Phase-I: training, PMixED follows the sample-and-aggregate paradigm \cite{nissim2007smooth} by partitioning a private corpus into $N$ pairwise disjoint subsets, then fine-tuning a pre-trained LLM on each subset to produce an ensemble. Using an ensemble is crucial for guaranteeing privacy, but also has been shown to boost perplexity \cite{jozefowicz2016exploring}. During Phase-II: private prediction, PMixED selects a subset of the ensemble using Poisson subsampling. Each selected model produces its output probability distribution, then follows the RD-mollification process by projecting its output distribution onto a Renyi Divergence ball centered at a public model's output distribution. Lastly, PMixED averages these optimal projections and then samples from it. 

Because PMixED does not employ differentially private training algorithms, it does not incur substantial training overhead like DP-SGD. Furthermore, we reduce computational and storage costs by employing parameter-efficient fine-tuning methods. In summary, our key contributions are the following: (1) We introduce and formalize $(\alpha, \beta)$-RD Mollifiers, a generalization of $\epsilon$-Mollifers, to avoid additive DP noise. (2) We propose a private prediction protocol utilizing RD Mollifiers and prove that it satisfies the DP prediction definition with group-level privacy. (3) We experimentally demonstrate that PMixED outperforms DP-SGD on two large-scale datasets. (4) We open-source our software implementation of PMixED to further spark research in this area\footnote{\url{https://github.com/james-flemings/pmixed}}.

\section{Related Works}

Most of the previous differential privacy work in LLMs has focused on private training. \citet{mcmahan2017learning} first explored private training of language models using a small recurrent neural network to achieve user-level differential privacy in the federated learning setting. Recent breakthroughs in differentially private LLMs involve self-supervised pre-training on public data, followed by privately fine-tuning on a private corpus \cite{li2021large, yu2021differentially}. 

An orthogonal approach, but conceptually similar to ours, is PATE \cite{papernot2016semi, papernot2018scalable}. PATE also uses an ensemble of models trained on pairwise disjoint subsets of a private dataset to generate DP labels. However, PATE and PMixED rely on substantially different private aggregation schemes. PATE uses the Gaussian/Laplacian mechanism to perturb its output vote count histogram while our method utilizes the inherent stochastic nature of sampling from a probability distribution and a public model in LLMs. Furthermore, PATE does not satisfy the DP prediction definition \cite{dwork2018privacy} since its data-dependent privacy accounting causes its privacy loss to be a function of the private data. Hence, testing that the data-dependent loss does not exceed the privacy budget will leak additional privacy \cite{redberg2023generalized}. 

There is a much smaller body of work that has focused on private prediction for generative language models, mainly due to prior work empirically showing that private prediction methods perform worse than private training \cite{van2020trade}. Private prediction can be broadly categorized into two methods: prediction sensitivity, which adds noise to the output distribution of an LLM; and sample-and-aggregate, which involves the same process as PATE for generating noisy labels. For prediction sensitivity, \citet{majmudar2022differentially} used the uniform distribution as their perturbation mechanism and showed that it satisfies $\epsilon$-DP. However, the privacy loss needs to be large, $\epsilon \approx 60$, in order to be practical.

One work closely related to PMixED, which motivated our work, also falls under the sample-and-aggregate method as ours \cite{ginart2022submix}. However, their ensemble could exceed the privacy budget before completing all $T$ queries due to their data-dependent privacy accounting. Hence, they had to define a new privacy notion to account for this. Our work does not have this limitation.

\section{Preliminaries}
\subsection{Next-Token Prediction Task}
Given some context vector $\mathbf{x}_t = x_1, x_2, ..., x_t$, which is a string of tokens from some vocabulary $V$, i.e. $x_i \in V $ for all $i=1, ..., t$, the task is to predict the next token $x_{t+1}$ using a generative language model $p$. More precisely, the output of a language model $p$ for a given context $\mathbf{x}_t$ is a likelihood function of all possible tokens $p(x_{t+1} = w | \mathbf{x}_t)$, and choosing the next token involves sampling from this probability mass function to obtain a token $\hat{x}_{t+1} \sim p(x_{t+1}| \mathbf{x}_t)$.

\subsection{Differential Privacy}
If a machine learning model has memorized any sensitive information that is contained in its training data, then it can potentially reveal this information during prediction. Differential Privacy (DP) is a mathematical framework that gives privacy guarantees for this type of privacy leakage by reducing the effect any individual has on a model. 

\begin{definition}[Approximate DP \cite{dwork2014algorithmic}] More formally, let $\epsilon>0, \delta \in [0, 1]$. A randomized algorithm $A: \mathcal{D} \rightarrow \mathcal{R}$ satisfies $(\epsilon, \delta)$-DP if for any pair of adjacent datasets $D, D^{'} \in \mathcal{D}$ and any set of subset of outputs $S \subseteq \mathcal{R}$ it holds that:
\begin{equation*}
    \Pr[A(D) \in S] \leq e^{\epsilon} \Pr[A(D^{'}) \in S] + \delta.
\end{equation*}
\end{definition}

The privacy parameters $\epsilon, \delta$ can be interpreted as follows: $\epsilon$ upper bounds the privacy loss, and $\delta$ is the probability that this guarantee does not hold. One subtle, but crucial, technicality is that adjacency can be achieved at any granularity. E.g., $D'$ adds or removes $k$ entries from $D$, which is known as the "add\textbackslash remove" scheme, or $D$ and $D'$ differ by $k$ entries where $0 < k \leq n$, which is known as the replacement scheme. Technically, both schemes are equivalent but for this work, we focus on the "add\textbackslash remove" scheme.



Renyi DP (RDP), another notion of DP, contains composition properties that are easier to work with than $(\epsilon, \delta)$-DP \cite{mironov2017renyi}. To define RDP, we first define Renyi Divergence. 

\begin{definition}[Renyi Divergence \cite{mironov2017renyi}]
    For two probability distributions $P$ and $Q$ defined over $\mathcal{R}$, the Renyi divergence of order $\alpha > 1$ is
    \begin{equation}
        D_{\alpha}(P || Q) = \frac{1}{\alpha-1} \log \mathop{\mathbb{E}}_{x \sim Q} \left[\left ( \frac{P(x)}{Q(x)} \right)^{\alpha} \right],
    \end{equation}
\end{definition}
\noindent and $D_{\alpha}^{\leftrightarrow}(P || Q) = \max\{D_{\alpha}(P || Q), D_{\alpha}(Q || P)\}$.

\begin{definition}[$(\alpha, \epsilon)$-RDP\cite{mironov2017renyi}]
    A randomized algorithm $A: \mathcal{D} \rightarrow \mathcal{R}$ is $(\alpha, \epsilon)$-RDP if for any adjacent datasets $D, D^{'} \in \mathcal{D}$ it holds that 
    \begin{equation}
        D_{\alpha}(A(D) || A(D^{'})) \leq \epsilon.
    \end{equation}
\end{definition}

RDP possesses useful properties which we will make use of in our privacy analysis and discuss in Appendix \ref{sec:useful_thms}.

\subsection{Private Training vs. Private Prediction}
We say that $A$ is a private training algorithm if it produces weights that are differentially private with respect to a private corpus $D$. For DP-SGD, $A(D)$ returns the per-sample gradients of the parameters perturbed with Gaussian noise for each iteration of training. Private training algorithms provide strong privacy since they limit privacy leakage even when an adversary has complete, white-box access to the model parameters.

We observe that commercial deployments of LLMs are typically cloud-based and the model is only accessible through API, so an adversary does not have access to model parameters. In this black box setting an alternative approach, called private prediction \cite{dwork2018privacy}, provides DP at prediction, which exploits this important relaxation. We formalize this below:

\begin{definition}[Privacy-Preserving Prediction]\label{def:pp-prediction}
   Let $A$ be a non-private training algorithm such that $p = A(D)$, $Q$ be an interactive query generating algorithm that generates queries $\mathbf{x}_t$, and $\mathcal{P}$ be a protocol that responds with $y_t \in V$. Define the output $(Q \rightleftharpoons_{T} \mathcal{P}(\theta)) = \{(\mathbf{x}_t, y_t)\}_{t=1}^{T}$ as a sequence query-response pairs where $T > 0$ is some positive integer. Then $\mathcal{P}$ is a private prediction protocol if $(Q \rightleftharpoons_{T} \mathcal{P}(\theta))$ is $(\alpha, \epsilon)$-RDP, i.e., for adjacent datasets $D, D^{'}$ with $p = A(D)$ and $p' = A(D^{'})$:
   \begin{equation*}
       D_{\alpha}((Q \rightleftharpoons_{T} \mathcal{P}(p) || (Q \rightleftharpoons_{T} \mathcal{P}(p'))) \leq \epsilon.
   \end{equation*}
\end{definition}
Note that due to the Post-Processing Theorem \ref{thm:post-processing}, any predictions made by a privately-trained model are differentially private. Alternatively, the goal of our work, given a private budget $\epsilon_G$, is to develop a protocol $\mathcal{P}$ that uses a non-private model $p$ to make $(\alpha, \epsilon_G)$-RDP predictions on a sequence of queries $\{\mathbf{x}_t\}_{t=1}^{T}$.
\section{PMixED: A Protocol For Private Next Token Prediction}\label{pmixed}
We now introduce and describe PMixED, a private prediction protocol for next token prediction. First, we will introduce a concept called Renyi Divergence (RD) Mollifiers, which formalize the projection of a private distribution onto a set around a public distribution, and help us prove the privacy guarantees of PMixED. Next, we will discuss the fine-tuning process of the ensemble, which follows the sample-and-aggregate paradigm. Lastly, we will describe the prediction protocol and show its privacy guarantees.

\subsection{Renyi Divergence Mollifiers}
Results from \citet{ginart2022submix, majmudar2022differentially} showed that perturbing the output probability distribution of an LLM with noise significantly degrades the performance. We take a different approach by leveraging the inherent probabilistic nature of next-token prediction, which involves sampling the output distribution of an LLM. Hence, the output distribution of each fine-tuned LLM is mixed with the output distribution of a public LLM, thereby reducing the memorization obtained by fine-tuning. More formally, a privacy-sensitive distribution $P$ is projected onto a given RD mollifier, which is a set of distributions such that the Renyi Divergence of any pair of distributions in this mollifier does not diverge by too much while preserving the modes of $P$. This is pictorially shown in figure \ref{fig:mollifier_projection}, and we formally define this below:

\begin{figure}
    \begin{minipage}[t]{0.45\columnwidth}
        \centering
        \includegraphics[width=\columnwidth]{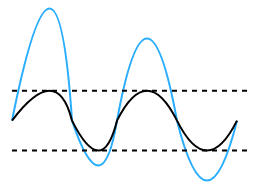}
    \end{minipage}
    \begin{minipage}[t]{0.45\columnwidth}
        \centering
        \includegraphics[width=0.9\columnwidth]{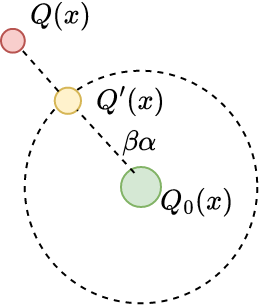}
    \end{minipage}
    \caption{\textit{Left:} Projecting a distribution $P$ (blue curve) onto an $(\alpha, \beta)$-RD mollifier (black curve). The dotted line represents the maximum divergence $\beta\alpha$ of the mollifier. Note how the projected distribution maintains the same modes as $P$. \textit{Right:} $Q'$ is the maximized projection of $Q$ onto a relative RD-mollifier around $Q_{0}$, which diverges by at most $\beta \alpha$ from $Q_{0}$.}
    \label{fig:mollifier_projection}
\end{figure}

\begin{definition}[$(\alpha, \beta)$-RD Mollifier] Let $\mathcal{M} \subset \mathcal{D}(\mathcal{X})$ be a set of distributions and $\alpha > 1, \beta > 0$. Then $\mathcal{M}$ is an $(\alpha, \beta)$-RD mollifier iff for all $Q, Q^{'} \in \mathcal{M}$, $x \in \mathcal{X}$
\begin{equation}\label{eq:rd-mollifier}
    D_{\alpha}(Q(x) || Q^{'}(x)) \leq \beta \alpha .
\end{equation}
\end{definition}
For example, the singleton $\mathcal{M}=\{Q\}$ is an $(\alpha, 0)$-RD mollifier. Note that an $\epsilon$-Mollifier from \cite{husain2020local} is also an $(\alpha, \frac{1}{2}\epsilon^2)$-RD mollifer for all $\alpha$ by the conversion from pure to zero concentrated DP \cite{bun2016concentrated}. RD-mollifiers consist of distributions that are close to each other with respect to the Renyi Divergence. \citet{husain2020local} states that deriving mollifiers is not always clear, but one way is to start from a reference distribution $Q_{0}$ and consider the set of all distributions that are close to $Q_{0}$, which we define below: 

\begin{definition}[$(\alpha, \beta)$-RD Mollifier relative to $Q_{0}$] Let $\mathcal{M}_{(\alpha, \beta), Q_{0}} \subset \mathcal{D}(\mathcal{X})$ be a set of distributions. Then for all $Q \in \mathcal{M}_{(\alpha, \beta), Q_{0}}$  
\begin{equation*}
    D_{\alpha}^{\leftrightarrow}(Q(x) || Q_{0}(x)) \leq \beta \alpha.
\end{equation*}
\end{definition}

\begin{lemma}\label{lemma:rel_conv}
    If $\mathcal{M}_{(\alpha, \beta), Q_{0}}$ is an $(\alpha, \beta)$-RD Mollifier relative to $Q_{0}$, then $\mathcal{M}_{(\alpha, \beta), Q_{0}}$ is an $(\alpha, 4\beta)$-RD Mollifer.
\end{lemma}
\textit{Proof}. A straightforward application of the Triangle-like inequality property of Renyi Divergence, Theorem \ref{thm:weak-triangle-inequality}. \qed

The goal then becomes taking a distribution $P$ and finding a distribution $\hat{P}$ inside a given mollifier $\mathcal{M}$ that minimizes the RD divergence:
\begin{equation} \label{eq:optimal_projection}
    \hat{P} \in \argmin_{Q\in \mathcal{M}} D_{\alpha}(P || Q).
\end{equation}

This process is called RD-mollification. In the following subsection, we will describe a mollification mechanism that takes a privacy-sensitive distribution $P$ as input and outputs a mollified distribution $\hat{P}$ that maximizes utility by finding the closest distribution to $P$ in a given mollifier.

\subsection{Fine-tuning a Pre-trained Model}
Suppose we have access to a non-private training procedure $A$, which takes as input a private dataset $D$ and a set of pre-trained weights $p_{0}$. Then $A(D, p_{0})$ returns a set of weights that have been fine-tuned on $p_{0}$ using $D$. Our instantiation of the fine-tuning process utilizes LoRA for parameter efficiency~\cite{hu2021lora}, however, any fine-tuning method can be used. First we partition a private dataset $D$ into $N$ subsets $D_{1}, D_{2}, ... D_{N}$ such that they are pairwise disjoint, i.e., $D_{i} \cap D_{j} = \varnothing$ for $i \neq j$, and $|D_{i}| = |D| / N$. Then for each subset $D_{i}$, we fine-tune $p_{0}$ using our training procedure $A$ to produce $p_{i} = A(D_{i}, p_{0})$. 

We want to highlight why LoRA is the natural parameter-efficient fine-tuning method to apply in this setting, and how it makes an ensemble of LLMs practical. Although PMixED conceptually produces $N$ models by the end of the fine-tuning process, using LoRA in our implementation only needs one model for inference, which is the pre-trained model $p_{0}$ combined with a set of LoRA adapter weights $p_{i}$. So when the $i$-th model performs inference, we just replace the LoRA weights with $p_{i}$ while still using the pre-trained model $p_{0}$. 

\subsection{Differentially Private Prediction Protocol}
Given a sequence of queries $\{\mathbf{x}_t\}_{t=1}^{T}$, PMixED responds to each query $\mathbf{x}_t$ in a differentially private manner. This is succinctly summarized in Algorithm \ref{alg:pmixed_protocol} and is broken down into three steps:

\begin{algorithm}[tb]
    \caption{PMixED: A protocol for Private Next Token Prediction}
    \label{alg:pmixed_protocol}
    \textbf{Input.} Number of LLMs $N$, Fine-Tuned LLM's $\{p_{i}\}_{i=1}^{N}$, public model $p_{0}$, total number of queries $T$, privacy budget $\epsilon_G > 0$, Renyi Divergence order $\alpha > 1$, subsample probability $0 < q \leq 1$, a series of queries $\{\mathbf{x}_t\}_{t=1}^{T}$, Subsampled privacy loss function $\epsilon'_q(\alpha)$
    \begin{algorithmic}[1]
        \For {$t=1, ..., T$}
            \State Select a subset $S_t \subseteq [N]$ by choosing each model with probability $q$.
            \State $\beta 
            \gets \argmax_{\beta'} \left\{\epsilon'_q(\alpha) \leq \epsilon_G / T \right\}$
            \For{$i \in S_t$}
                \State $\lambda_i \gets$ using eq. \ref{eq:solve-lambda}
                \State $\overline{p}_{i}(\mathbf{x}_t) = \lambda_i p_{i}(\mathbf{x}_t) + (1-\lambda_i) p_{0}(\mathbf{x}_t)$ 
            \EndFor
            \State $p(\mathbf{x}_t) = p_{0}(\mathbf{x}_t)$
            \If {$S_t \neq \emptyset$}
                \State $p(\mathbf{x}_t) = \frac{1}{|S_t|} \sum\limits_{i \in S_t} \overline{p}_i(\mathbf{x}_t)$ 
            \EndIf
            \State $y_{t} \sim p(\mathbf{x}_t)$ 
        \EndFor
        \State \Return $\{y_1, ..., y_T\}$
    \end{algorithmic}
\end{algorithm} 

\begin{description}[style=unboxed, leftmargin=0cm]

\item[Poisson Subsampling of Ensemble.] We perform Poisson subsampling on the entire ensemble $\{p_{i}\}_{i=1}^{N}$ to obtain a subset of the ensemble $S_{t} \subseteq [N]$ such that each model $p_{i}$ is selected with probability $q$. The benefit of subsampling a subset of the ensemble is that it can further amplify the privacy of our protocol, since running on some random subset of the ensemble introduces additional uncertainty. In particular, if a protocol is $(\epsilon, \delta)$-DP then with subsampling probability $q$ it is roughly $(O(q\epsilon), q\delta)$-DP \cite{steinke2022composition}. In the case that no models are sampled, PMixED resorts to the public model $p_{0}$ for prediction entirely.

\item[Inference and RD-Mollification.] Each sampled model $i \in S_t$ performs inference to produce an output distribution $p_{i}(\mathbf{x}_t)$. Then we RD-mollify each distribution $p_{i}(\mathbf{x}_t)$ by mixing it with the public distribution $p_{0}(\mathbf{x}_t)$ using a mixing parameter $\lambda_i$ to produce $\overline{p}_{i}(\mathbf{x}_t) = \lambda_i p_{i}(\mathbf{x}_t) + (1-\lambda_i)p_{0}(\mathbf{x}_t)$. $\lambda_i$ is automatically chosen by solving the following optimization scheme:
\begin{equation}\label{eq:solve-lambda}
   \lambda_i \gets \argmax\limits_{\lambda \in [0,1]} \{ D_{\alpha}^{\leftrightarrow} (\overline{p}_{i}(\mathbf{x}_t) || p_{0}(\mathbf{x}_t)) \leq \beta\alpha \}.
\end{equation}
The value of $\beta$ will be specified once we state the privacy guarantees of our protocol. We opt to numerically solve Equation \ref{eq:solve-lambda} by using the bisection method from the \texttt{SciPy} library. Note that as we increase $\lambda_i$, the function $D_{\alpha}^{\leftrightarrow} (\overline{p}_{i}(\mathbf{x}_t) || p_{0}(\mathbf{x}_t))$ will increase because $\overline{p}_{i}(\mathbf{x}_t)$ diverges more from $p_{0}(\mathbf{x}_t)$ and approaches $p_{i}(\mathbf{x}_t)$. Hence $D_{\alpha}^{\leftrightarrow} (\overline{p}_{i}(\mathbf{x}_t) || p_{0}(\mathbf{x}_t))$ is monotonically increasing with respect to $\lambda_i$, which allows us to use bisection for this function. Moreover, each of the projected distributions is an element in the RD-Mollifer relative to $p_{0}(\mathbf{x}_t)$, i.e., $\overline{p}_{i}(\mathbf{x}_t) \in \mathcal{M}_{(\alpha, \beta), p_{0}(\mathbf{x}_t)}$ for all $i \in [N]$, which satisfies Equation \ref{eq:optimal_projection}, giving us the optimal projection.
\item[Aggregation and Sampling.] The last step is to average the projected distributions, then sample from this averaged distribution: \\ $y_{t} \sim \frac{1}{|S_t|} \sum\limits_{i \in S_t} \lambda_i p_{i}(\mathbf{x}_t) + (1-\lambda_i) p_{0}(\mathbf{x}_t)$. 

\end{description}
\section{Privacy Analysis}

We begin our privacy analysis of PMixED by first considering the case of no Poisson subsampling. Then we invoke the privacy amplification theorem \ref{thm:priacy_amp} to derive our final privacy guarantees. Lastly, we will discuss the implications of our privacy guarantees. Let $\mathcal{P}$ denote our private prediction protocol, PMixED. Note that $D$ and $D^{'}$ are neighboring datasets if $D^{'}$ adds or removes a subset $D_{i}$ from $D$, which is equivalent to adding or removing the model $p_i$ from the ensemble.   

\subsection{DP Guarantees for PMixED}
At a high level, we will prove that $\mathcal{P}$ is $(\alpha, \epsilon_G / T)$-RDP for query $\mathbf{x}_t$. Then, using the Composition Theorem \ref{thm:composition}, we can say that $\mathcal{P}$ is $(\alpha, \epsilon_G)$-RDP for query-responses $\{(\mathbf{x}_t, y_t)\}_{t=1}^{T}$, thus satisfying the Private Prediction Definition \ref{def:pp-prediction}. Appendix \ref{sec:intuition} analyzes the case when $\alpha = \infty$, which is pure DP. 
\begin{theorem}\label{prop:pmixed_priv_guarantee}
    Let 
   \begin{equation}\label{eq:beta_renyi}
   \beta \leq \begin{cases}
   \frac{\log \left (Ne^{(\alpha-1) \epsilon_G / T} + 1 - N \right)}{4 (\alpha - 1)\alpha}, &\text{if } N > 1 \\
   \frac{\epsilon_G}{T\alpha} &\text{otherwise}
   \end{cases}.
   \end{equation}
    Then the output of PMixED $\mathcal{P}$ on query $\mathbf{x}_t$ is $(\alpha, \epsilon_G / T)$-RDP with respect to $D$.
\end{theorem}

\noindent \textit{Proof.} Let $i \in [N]$ and $\mathbf{x}_t$ be a query. Define 
\begin{align*}
    &p(\mathbf{x}_t) = \frac{1}{N} \sum_{i=1}^{N} \lambda_i p_{i}(\mathbf{x}_t) + (1-\lambda_i) p_{0}(\mathbf{x}_t), \\
    &p_{-i}(\mathbf{x}_t) = \frac{1}{N-1} \sum_{j\neq i} \lambda_j p_{j}(\mathbf{x}_t) + (1-\lambda_j) p_{0}(\mathbf{x}_t)
\end{align*}
 where $\lambda_i$ is selected from Equation \ref{eq:solve-lambda}. Now, observe that each $\lambda_i$ is dependent only on $D_i$, so $p_{-i}(\mathbf{x}_t)$ does not contain $D_i$. Using the fact that $D_{\alpha}^{\leftrightarrow}(\overline{p}_j(\mathbf{x}_t) || p_{0}(\mathbf{x}_t)) \leq \beta \alpha$ for all $j\in[N]$, then for any two neighboring ensembles $\{p_i\}_{i=1}^{N}$, $\{p_{j}\}_{j\neq i}$ with $N > 1$
\begin{align}
    &e^{(\alpha-1) D_{\alpha}\left(y_t \sim \mathcal{P}\left(\{p_i\}_{i=1}^{N}, \mathbf{x}_t \right) || y_t \sim \mathcal{P}\left(\{p_j\}_{j\neq i}, \mathbf{x}_t \right)\right)} \nonumber \\
    &=e^{(\alpha-1) D_{\alpha}(p(\mathbf{x}_t) || p_{-i}(\mathbf{x}_t))} \nonumber \\
    &= \mathop{\mathbb{E}}_{p_{-i}(\mathbf{x}_t)} \left [ \left (\frac{p(\mathbf{x}_t)}{p_{-i}(\mathbf{x}_t)} \right )^{\alpha} \right] \nonumber \\
    &= \mathop{\mathbb{E}}_{p_{-i}(\mathbf{x}_t)} \left [ \left (\frac{\frac{N-1}{N}p_{-i}(\mathbf{x}_t) + \frac{1}{N}\overline{p}_{i}(\mathbf{x}_t)}{p_{-i}(\mathbf{x}_t)} \right )^{\alpha} \right] \nonumber \\
    &\leq \mathop{\mathbb{E}}_{p_{-i}(\mathbf{x}_t)} \left [ \frac{\frac{N-1}{N}(p_{-i}(\mathbf{x}_t))^{\alpha} + \frac{1}{N} (\overline{p}_{i}(\mathbf{x}_t))^{\alpha}}{(p_{-i}(\mathbf{x}_t))^{\alpha}} \right] \label{jensens} \\
    &= \mathop{\mathbb{E}}_{p_{-i}(\mathbf{x}_t)} \left [\frac{N-1}{N} + \frac{1}{N} \left( \frac{\overline{p}_{i}(\mathbf{x}_t)}{p_{-i}(\mathbf{x}_t)} \right)^{\alpha} \right] \nonumber \\
    &= \frac{N-1}{N} + \frac{1}{N} \mathop{\mathbb{E}}_{p_{-i}(\mathbf{x}_t)} \left [\left( \frac{\overline{p}_{i}(\mathbf{x}_t)}{p_{-i}(\mathbf{x}_t)} \right)^{\alpha} \right] \nonumber \\
    &= \frac{N-1}{N} + \frac{1}{N}e^{(\alpha-1)D_{\alpha}(\overline{p}_{i}(\mathbf{x}_t) || p_{-i}(\mathbf{x}_t))} \nonumber \\
    &\leq \frac{N-1}{N} + \frac{1}{N}e^{(\alpha-1)4\beta\alpha} \label{eq:quasi}
\end{align}
\noindent where Equation \ref{jensens} uses Jensen's inequality for the convex function $f(x) = x^{\alpha}$ since $\alpha \geq 1$ and $x \geq 0$ because we are dealing with probabilities, and Equation \ref{eq:quasi} is due to $D_{\alpha}(\overline{p}_{i}(\mathbf{x}_t) || p_{0}(\mathbf{x}_t)) \leq \beta\alpha$ and 
\begin{align*}
    D_{\alpha}(p_{0}(\mathbf{x}_t) || p_{-i}(\mathbf{x}_t)) 
    &\leq \max_{j \neq i}  D_{\alpha}(p_{0}(\mathbf{x}_t) || \overline{p}_{j}(\mathbf{x}_t)) \\
    &\leq \beta \alpha
\end{align*}
 by the Quasi Convexity property of Renyi Divergence \ref{thm:quasi-convexity}. Then using the Triangle-like Inquality \ref{thm:weak-triangle-inequality} gives us $D_{\alpha}(\overline{p}_{i}(\mathbf{x}_t) || p_{-i}(\mathbf{x}_t)) \leq 4 \beta \alpha$. Hence plugging in Equation \ref{eq:beta_renyi} for $\beta$ into Equation \ref{eq:quasi} implies $D_{\alpha}(p(\mathbf{x}_t) || p_{-i}(\mathbf{x}_t)) \leq \epsilon_G / T$. When $N=1$, then $p_{-i}(\mathbf{x}_t) = p_{0}(\mathbf{x}_t)$ since our protocol will resort to using $p_{0}$. 

Now, for the other way
\begin{align}
    &D_{\alpha}(p_{-i}(\mathbf{x}_t) || p(\mathbf{x}_t)) \nonumber \\
    &=\textstyle D_{\alpha}\left(p_{-i}(\mathbf{x}_t) \Big\| \frac{N-1}{N}p_{-i}(\mathbf{x}_t) + \frac{1}{N}\overline{p}(\mathbf{x}_t)\right)\nonumber \\
    &\textstyle \leq \frac{N-1}{N}D_{\alpha}(p_{-i}(\mathbf{x}_t) || p_{-i}(\mathbf{x}_t))\nonumber \\ 
    &+ \textstyle\frac{1}{N} D_{\alpha}(p_{-i}(\mathbf{x}_t) || \overline{p}_{i}(\mathbf{x}_t))\label{eq:sec-convex} \\
    &\leq \frac{4\beta\alpha}{N}\label{eq:converse-quasi} \\
    & = \frac{\frac{N-1}{N}\log(1) + \frac{1}{N}\log\left(\exp\left(4\beta\alpha(\alpha-1)\right)\right)}{\alpha-1}\nonumber \\
    &\leq \frac{\log\left(\frac{N-1 + \exp\left({(\alpha-1)4\beta\alpha}\right)}{N} \right)}{\alpha-1} \label{eq:beta}
\end{align}
where Equation \ref{eq:sec-convex} uses convexity in the second argument of Renyi divergence (Theorem \ref{thm:second-convexity}), Equation \ref{eq:converse-quasi} uses the same argument as in Equation \ref{eq:quasi}, and Equation \ref{eq:beta} is due to concavity of logarithms. Then setting $\beta$ to Equation \ref{eq:beta_renyi} gives us $D_{\alpha}(p_{-i}(\mathbf{x}_t) || p_(\mathbf{x}_t)) \leq \epsilon_G / T$. Thus proves the claim that the output of $\mathcal{P}$ on query $\mathbf{x}_t$ is $(\alpha, \epsilon_G / T)$-RDP. \qed

\begin{theorem}
    PMixED $\mathcal{P}$ is an $(\alpha, \epsilon_G)$-RDP prediction protocol with respect to $D$.
\end{theorem}

\noindent \textit{Proof.} Let $Q$ be an interactive query generating algorithm that generates queries $\mathbf{x}_t$. We first obtain fine-tuned weights using our training algorithm $p_i = A(p_{0}, D_i)$. Then $\mathcal{P}$ uses $\mathbf{x}_t$ as input and returns a response, which is a sample $y_t \sim \mathcal{P}(\{p_i\}_{i=1}^{N}, \mathbf{x}_t)$. By Theorem \ref{prop:pmixed_priv_guarantee}, $y_t$ is $(\alpha, \epsilon_G / T)$-RDP. Then after $T$ queries and responses, the sequence $\{(\mathbf{x}_t, y_t)\}_{t=1}^{T}$ is $(\alpha, \epsilon_G)$-RDP by the Composition Theorem \ref{thm:post-processing}. Therefore $\mathcal{P}$ is an $(\alpha, \epsilon_G)$-RDP prediction protocol. \qed

A few observations to highlight: (1) $\beta$ depends on the number of models in the ensemble $N$. As $N$ increases, then $\beta$ also becomes larger. Intuitively, each model has less effect on the overall output, which allows them to diverge more from the public model. Also note that the choice of $N$ is fixed by the analyst, independent of the dataset. Hence $\beta$ does not leak information about the dataset. 

(2) Our analysis also relies on the fact that PMixED employs ancestral sampling as its decoding strategy, which samples directly from the ensemble's per query distribution, $p(\mathbf{x}_t)$. Truncated decoding such as top-k \cite{fan2018hierarchical} or top-p \cite{holtzman2019curious} sampling which samples only plausible tokens in the distribution, and greedy decoding which only samples the most likely next token, could be employed to improve the text generation quality. However, additional privacy leakage can occur, so we leave it as a future work to extend PMixED to these decoding strategies.  

\subsection{DP Guarantees for Subsampled PMixED}
Now that we have shown that $\mathcal{P}$ is an $(\alpha, \epsilon_G)$-RDP prediction protocol with the "add\textbackslash remove" scheme, PMixED is compatible with the Privacy Amplification by Poisson Subsampling Theorem \ref{thm:priacy_amp}. Let $S_{t}$ be the subsampled set at time $t$ where $p_{S_t}$ is $(p_{S_t})_i = p_i$ if $i \in S_t$ and $(p_{S_t})_i = \bot$ if $i \notin S_t$. Let $\mathcal{P}^{S_t}$ be the subsampled PMixED protocol where $\mathcal{P}^{S_t}(p, \mathbf{x}_t) = \mathcal{P}(p_{S_t}, \mathbf{x}_t)$. Then by Theorem \ref{thm:priacy_amp}, the output of $\mathcal{P}^{S_t}$ on query $\mathbf{x}_t$ is $(\alpha, \epsilon'_{q}(\alpha))$-RDP where $\epsilon'_{q}(\alpha)$ is defined in Equation \ref{eq:subsam_eps}.

Since $\epsilon^{'}_{q}(\alpha) \ll \epsilon_{G} / T$, we can increase $\epsilon^{'}_{q}(\alpha)$ to be as close to $\epsilon_G / T$ as possible by increasing $\epsilon(\alpha)$, the privacy loss of $\mathcal{P}$, using $\beta$. In other words, the privacy loss of PMixED is
\[
   \epsilon(\alpha) \leq
   \begin{cases}
   \frac{\log\left(\frac{|S_t|-1 + \exp\left({(\alpha-1)4\beta\alpha}\right)}{|S_t|} \right)}{\alpha-1} &\text{if } |S_t| > 1 \\
   \beta \alpha &\text{otherwise}
   \end{cases}.
\]
 Hence, PMixED uses $\epsilon(\alpha)$ to solve the following equation: $\beta^{*} = \argmax_{\beta} \left\{\epsilon'_q(\alpha) \leq \epsilon_G / T \right\}$, which selects the optimal RD radius $\beta^{*}$.

\subsection{Privacy Level Granularity}
Seemingly, PMixED is stronger than sample level privacy since $D_{\alpha}^{\leftrightarrow}(\mathcal{P}(D) || \mathcal{P}(D \setminus \{d\})) \leq D_{\alpha}^{\leftrightarrow}(\mathcal{P}(D) || \mathcal{P}(D \setminus D_i))$ for some sample $d \in D_i$ and $i \in [N]$, due to the group privacy property \cite{mironov2017renyi}. In actuality, our method is closely related to group-level privacy \cite{ponomareva2023dp}, where each subset $D_i$ defines a group of samples, and each sample is contained in exactly one group. This flexibility allows PMixED to offer different granularities of privacy, depending on the partitioning of the private corpus. For example, PMixED is compatible with virtual client-level privacy \cite{Xu_2023_CVPR}, a stronger version of user-level privacy, where a subset $D_{i}$ is considered a virtual client comprised of groups of user data, and each user's data is stored in at most one subset. For practical language modeling datasets where users can contain multiple data samples, guaranteeing sample-level privacy is insufficient to ensure the privacy of an individual user \cite{mcmahan2017learning}. Hence, for these types of datasets, PMixED can provide a strong enough privacy guarantee for users. We delegate dataset partitioning, consequently determining the privacy level, to the practitioner.
\section{Experiments}

\begin{table}
\begin{center}
\begin{tabular}{||c c ||} 
     \hline
     Parameter & Value \\ [0.5ex] 
     \hline\hline
     Privacy Budget: $\epsilon_G$ & 8 \\ 
     \hline
     Runs:  &$32$ \\ 
     \hline
     Probability of Failure: $\delta$ & 1e-5 \\
     \hline
     Renyi Divergence Order: $\alpha$ & 3 \\
     \hline
     Inference Budget: $T$ & 1024 \\
     \hline
     Number of Ensembles: $N$ & 80 \\
     \hline
     Subsample Probability: $p$ & 0.03 \\ [1ex] 
     \hline
\end{tabular}
\caption{Privacy Hyperparameters for PMixED and DP-SGD.}
\label{table:pred_parameters}
\end{center}
\end{table}

\subsection{Experimental Setup}
We experimentally evaluated the privacy-utility tradeoff of PMixED by using LoRA \cite{hu2021lora} to fine-tune pre-trained GPT-2 models \cite{radford2019language} from HuggingFace \cite{wolf2019huggingface} on the WikiText-103 \cite{merity2016pointer} and One Billion Word \cite{chelba2013one} datasets. We view these two large word-level English language modeling benchmarks as complementary to each other, in that they give a good comprehensive evaluation of differentially private next token prediction. WikiText-103 tests the ability of long-term dependency modeling, while One Billion Word mainly tests the ability to model only short-term dependency \cite{dai2019transformer}. 

The public model in our experiments is a pre-trained GPT-2 small model. We compare PMixED to three baselines: the public model, a non-private fine-tuned model, and a private fine-tuned model produced by DP-SGD. Although DP-SGD has a weaker privacy level, we compared PMixED to per-sample DP-SGD as our baseline because it illustrates how well PMixED performs against the most widely-used DP solution. The non-private and private fine-tuned baselines also used LoRA. LoRA based  fine-tuning can outperform full private fine-tuning since the LoRA updates a much smaller set of parameters during fine-tuning\cite{yu2021differentially}. The LoRA parameters used for each model $p_i$ contain $0.11\%$ of the total number of parameters of the pre-trained model, allowing us to fit the entire ensemble into one GPU during prediction. Appendix \ref{sec:add_experiments} contains more details about fine-tuning.

For private prediction, unless stated otherwise, the parameters used are set to the default values shown in Table \ref{table:pred_parameters}. The privacy hyperparameters are reported in terms of $(\epsilon_G, \delta)$-DP, however we perform our privacy loss calculations in terms of RDP, then convert back using Theorem \ref{thm:rdp-dp}. To measure the utility, we use test-set perplexity with a sequence length of $512$. In total, $T$ predictions are made for each run, and a total of $32$ runs are performed and averaged for each baseline.

\subsection{Comparison Over Baselines}
\begin{figure}
    \centering
    \includegraphics[width=\columnwidth]{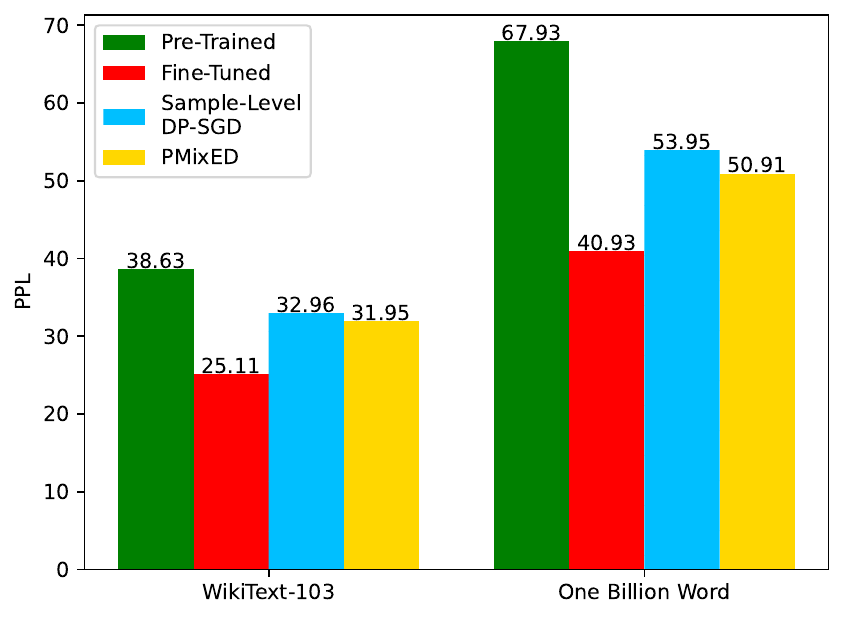}
    \caption{Comparison of PMixED against 3 baselines on  WikiText-103 and One Billion Word using GPT-2.}
    \label{fig:comparison}
\end{figure}

The pre-trained and fine-tuned models represent two extremes in the privacy-utility spectrum: the pre-trained model is perfectly private but has no utility gain, while the fine-tuned model has the best utility but guarantees no privacy. Since PMixED is a mixture of both, its utility and privacy guarantees should be between both. Figure \ref{fig:comparison}, which contains results of each baseline and PMixED on the test-set perplexity, shows that this is indeed the case. For the WikiText-103 dataset, the pre-trained model achieved a perplexity score of $38.63$, and the fine-tuned model achieved $25.11$. PMixED scored $31.95$ which is a $7$ point perplexity improvement over the pre-trained model, while guaranteeing $(\epsilon_G, \delta)$-DP as opposed to the completely non-private fine-tuned model $\epsilon_G=\infty$. Furthermore, PMixED gains a $17$ point perplexity improvement over the pre-trained model for the One Billion Word dataset. 

PMixED was also able to improve the perplexity score over DP-SGD by $1$ and $3$ points on the WikiText-103 and One Billion Word datasets, respectively. This result validates that private prediction methods can outperform private training for large query budgets \cite{van2020trade}. Moreover, the results of PMixED demonstrate that we can obtain the stronger group-level privacy without compromising utility. 

\subsection{Ablation Study}
\begin{figure}
    \begin{subfigure}{0.495\columnwidth}
        \centering
        \includegraphics[width=\columnwidth]{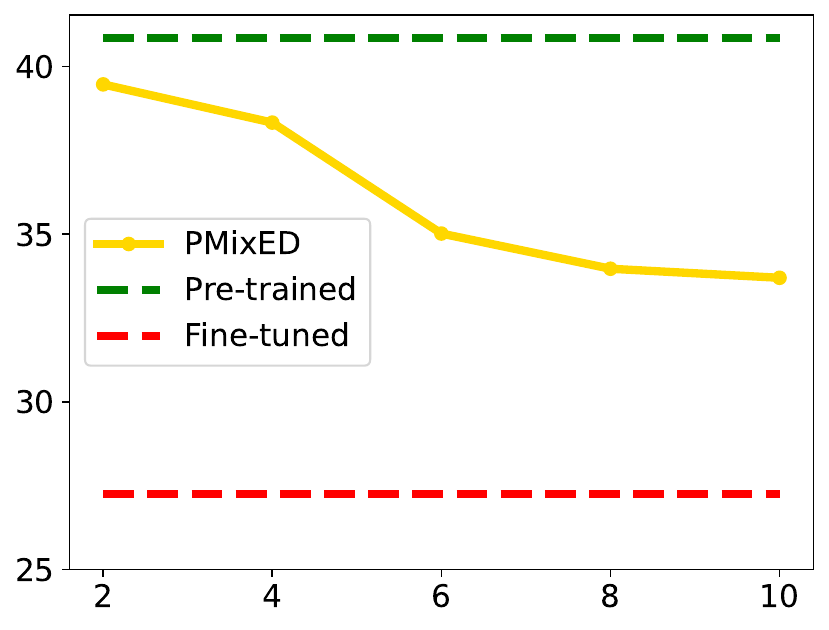}
        \caption{\small epsilon}
        \label{fig:epsilon}
    \end{subfigure}
    \hfill
    \begin{subfigure}{0.495\columnwidth}
        \centering
        \includegraphics[width=\columnwidth]{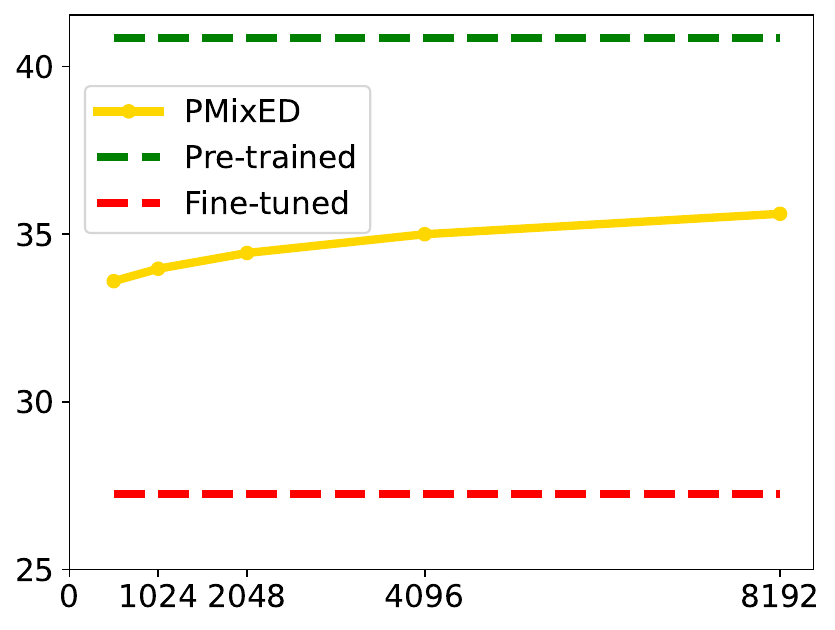}
        \caption{query budget}
        \label{fig:query_budget}
    \end{subfigure}
    \vskip\baselineskip
    \begin{subfigure}{0.495\columnwidth}
        \centering
        \includegraphics[width=\columnwidth]{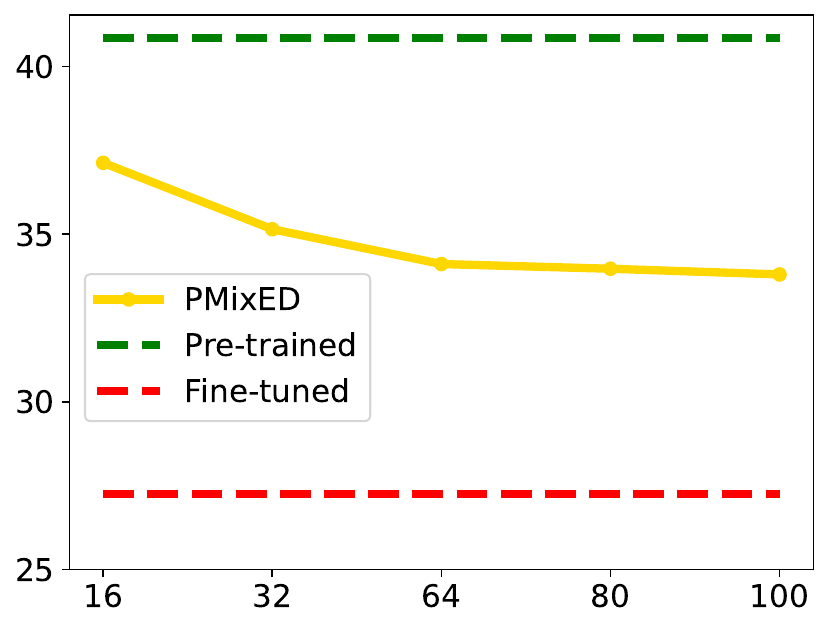}
        \caption{ensemble}
        \label{fig:ensemble}
    \end{subfigure}
    \hfill
    \begin{subfigure}{0.495\columnwidth}
        \centering
        \includegraphics[width=\columnwidth]{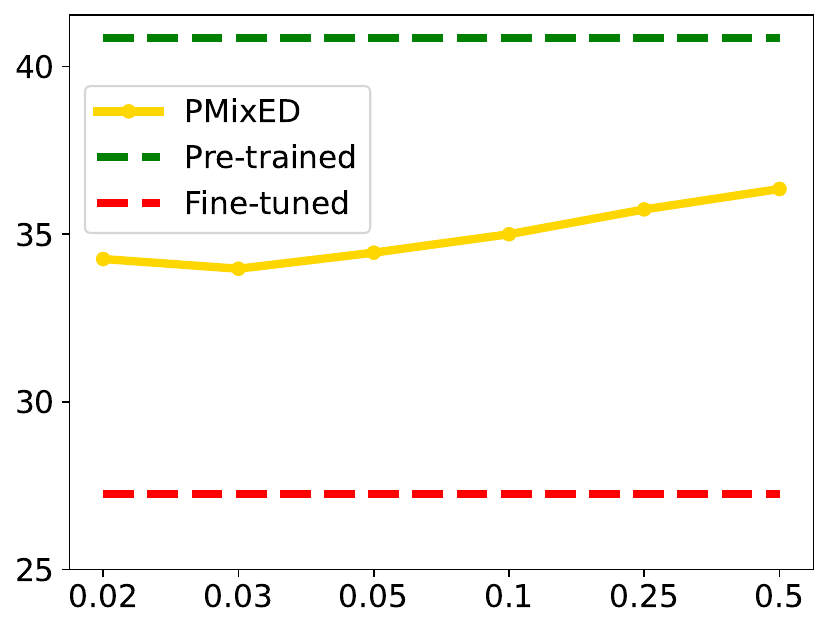}
        \caption{probability}
        \label{fig:probability}
    \end{subfigure}
    \caption{Ablation study on DP hyperparameters using WikiText-103. The x-axis is the hyperparameter space, and the y-axis is the perplexity score.}
    \label{fig:ablation study}
\end{figure}

We explore the privacy hyperparameters used by PMixED for prediction on WikiText-103. $8$ runs are performed and averaged for each hyperparameter value, shown in Figure \ref{fig:ablation study}.

Figure \ref{fig:epsilon} shows that for small privacy budgets $\epsilon_G \in [2, 4]$, the utility of PMixED approaches the pre-trained model. This is due to the RD mollification producing projected distributions close to the public model because $\epsilon_G$ is too small, so $\alpha$ must be large causing $D_{\alpha}$ to decrease monotonically. For moderately sized privacy budgets $\epsilon_G \in [4, 8]$, we see a sharp improvement in perplexity since $\alpha$ is smaller. Appendix \ref{sec:sup} shows the relationship between $\alpha$ and perplexity. Figure \ref{fig:query_budget} shows that even if the inference budget $T$ is as large as $T=8192$, we observe only marginal performance degradation of PMixED, losing only $1.6$ PPL from $T=1024$. Thus PMixED is capable of handling large amounts of inference while still being performant. 

For the results in Figure \ref{fig:ensemble}, ensemble sizes $N=16, 32, 64$ were trained with $E=10$ epochs, while $N=80$ was trained with $E=15$ and $N=100$ with $E=20$. The trend seems to be that larger ensembles with increased training epochs lead to better performance. We surmise that larger ensembles increase the expected number of subsampled models. And more explicit memorization occurs with increasing epochs, which when mixed with the public model allows for better generalization, similar to \cite{khandelwal2019generalization}. Eventually, too many models can lead to little training data to learn from, which we leave as future work to explore. 

 We conclude by remarking that in comparison to DP-SGD, the additional hyperparameters introduced by PMixED are the query budget $T$, the ensemble size $N$, and the subsample probability $q$. However, DP-SGD contains the clipping threshold and the expected subsample batch size as hyperparameters that PMixED does not need to work with. Therefore, since $T$ is determined based on the problem setup, PMixED does not add more hyperparameters than DP-SGD. Moreover, tuning the hyperparameters of PMixED is considerably faster compared to DP-SGD because $N$ and $q$ are tuned during prediction, as opposed to retraining an LLM for the tuning process of DP-SGD.
\section{Conclusion}
PMixED is inspired by the observation that most LLM deployments are cloud-based where an adversary only has black-box access to the model, rather than access to the entire model parameters. Under this setting, PMixED presents a novel private prediction protocol that provides DP during prediction, rather than during training. Thus PMixED is model agnostic, which has significant practical implications since it avoids the complexity of privately training models with millions or even billions of parameters. Our approach relies on the novel idea of $(\alpha, \beta)$-RD Mollifiers to project each of the private model's output distribution onto a set around a public LLM's output distribution. PMixED achieves the strong group-level DP and outperforms DP-SGD in terms of utility on large-scale datasets using LLMs. Thus, PMixED substantially progresses private prediction methods in LLMs and offers a practical alternative to DP training methods.
\section{Limitations}
PMixED performs better with larger ensemble sizes combined with longer training epochs, as quantified in the ablation studies. As ensemble sizes grow each model in the ensemble needs sufficient data to train, which in turn increases the cumulative training data size. We also note that PMixED requires more training epochs than regular SGD. However, these limitations are generally a bottleneck for all DP-based approaches, such as DP-SGD, and hence are not unique limitations of PMixED.

In this work, we employed LoRA to significantly reduce model size demands. However, storing an ensemble of models in memory can place a burden on the computing resources during training and inference procedures. The inference latency can be negatively impacted as an ensemble of models must provide predictions first followed by calculating the lambdas, due to the use of the bisection method. Potential systems optimizations can be to parallelize the inference and lambda calculations. We leave it as a future work to reduce the inference latency. But on the positive side, PMixED operates well with batch-based training and hence can take advantage of parallel GPU resources, while DP-SGD needs per-sample gradients which can curtail GPU efficiency.  

Perhaps the biggest limitation of PMixED is the query budget, which limits the number of predictions that can be made. However, this limitation is a natural consequence of differentially private prediction from a nonprivately trained model. As a future work, one can explore a more relaxed problem setup, a PATE-like setting \cite{papernot2016semi, papernot2018scalable}, where we produce DP predictions while minimizing the privacy loss. 

Finally, we note how work in DP-SGD, as well as this work, uses a pre-trained model to boost performance. Unfortunately, using publicly available data is not necessarily risk-free in terms of privacy, as prior works were able to extract personally identifiable information from a GPT-2 model pre-trained on data scraped from the public internet \cite{carlini2021extracting}. Thus model deployments must still be cautious of using pre-trained models in terms of understanding their information leakage potential.   
\section{Ethical Considerations}
In this work, we utilized pre-trained large language models and well-known language modeling datasets that were accessed from the Hugging Face API, which are publicly available and free to use. The GPT-2 model and the One Billion Word Dataset are licensed under the Apache License, Version 2.0, and the WikiText-103 dataset is licensed under CC BY-SA 3.0. Our intended use of these artifacts is aligned with the intended use of the creators, which, for us, is purely for benchmarking our private next token prediction protocol, and does not trademark these artifacts. Since the datasets were originally obtained from public internet domains, and seemingly do not contain personally identifiable information, we did not anonymize the dataset. Our use of public models and datasets minimizes any unintended privacy leakage that could result from experimenting. 

Additionally, we publicly released our code under the Apache License, Version 2.0. We believe that allowing open access to these experiments will help spark academic research of this work, as well as protect the privacy of user data in commercial deployment of large language models. 
\section*{Acknowledgements}
The authors would like to thank Tony Ginart for his tremendous advice and guidance early on in the project. This work is supported by Defense Advanced Research Projects Agency (DARPA) under Contract Nos. HR001120C0088, NSF award number  2224319, DGE-1842487, REAL@USC-Meta center, and gifts from Google and VMware. The views, opinions, and/or findings expressed are those of the author(s) and should not be interpreted as representing the official views or policies of the Department of Defense or the U.S. Government.


\bibliography{main}

\appendix
\newpage
\section{Additional Experimental Details}\label{sec:add_experiments}
The WikiText-103 dataset is a collection of over 100 million tokens from the set of verified Good and Bad articles on Wikipedia \cite{merity2016pointer}. The One Billion Word dataset is text data obtained from the Sixth Workshop on Machine Translation, and it contains nearly one billion words in the training set \cite{chelba2013one}. We split the datasets into sequences of length 512 tokens. A total of 920,344 examples are in the training set and 1928 in the validation set for WikiText-103. Table \ref{table:train_parameters} displays the hyperparameter values used for fine-tuning. Certain hyperparameter values for non-private fine-tuning were selected from \cite{hu2021lora}, and certain hyperparameter values for private fine-tuning from \cite{yu2021differentially}. We employed the AdamW optimizer with weight decay $0.01$ and a linear learning rate scheduler. Training ran on 8 Quadro RTX 5000, totaling around 8 hours for non-private fine-tuning, and 12 hours for private fine-tuning for WikiText-103. Prediction used only 1 Quadro RTX 5000.

There are technical challenges to get DP-SGD to contain the same privacy notion as ours. Either we would have to directly scale from sample-level privacy by the size of the partition $|D| / N$ using the group privacy property \cite{dwork2014algorithmic}, which incurs a prohibitive privacy cost. Or directly clip and add noise to the per-subset gradients, which is not possible to implement with standard DP libraries, like Opacus. 

\begin{table}
\begin{center}
\resizebox{\linewidth}{!}{%
\begin{tabular}{||c c c c||} 
     \hline
     Parameter & Fine-Tune & PMixED & DP-SGD\\ [0.5ex] 
     \hline\hline
     Epochs & 3 & 15\textsuperscript{\rm 1}, 5\textsuperscript{\rm 2} & 20\textsuperscript{\rm 1}, 9\textsuperscript{\rm 2} \\ 
     \hline
     Learning Rate & 2e-4 & 2e-4 & 4e-4\\
     \hline
     Weight Decay & 0.01 & 0.01 & 0.01\\
     \hline
     Adaptation $r$ & 4 & 4 & 4\\
     \hline
     LoRA $\alpha$ & 32 & 32 & 32\\
     \hline
     DP Batch Size & - & - & 256\\  
     \hline
     Clipping Norm & - & - & 1.0 \\ [1ex] 
     \hline
\end{tabular}}
\caption{Training Hyperparameters used for fine-tuning on \textsuperscript{\rm 1}WikiText-103 and \textsuperscript{\rm 2}One Billion Word.}
\label{table:train_parameters}
\end{center}
\end{table}

\section{Preliminary Setup for Theorem \ref{prop:pmixed_priv_guarantee}}
\label{sec:intuition}

Using our RD-mollifers concept, we introduce an additional concept called $(\alpha, \beta)$-RD private samplers, which is just $(\alpha, \beta)$-RDP but the neighboring condition is at the distributional level, then state how RD-mollifiers relates to RD-private samplers

\begin{definition}[$(\alpha, \beta)$-RD private sampler]
  Let $\alpha > 1, \epsilon > 0$. An $(\alpha, \beta)$-RDP sampler is a randomized mapping $M : \mathcal{D}(\mathcal{X}) \rightarrow \mathcal{X}$ such that for any $x \in \mathcal{X}$ and any two distributions $P, P^{'} \in \mathcal{D}(\mathcal{X})$ we have
   \begin{equation}
       D_{\alpha}(M(P) || M(P')) \leq \beta\alpha.
   \end{equation}
\end{definition}

\begin{lemma}\label{mollifier-sampler}
    Let $A: \mathcal{D}(\mathcal{X}) \rightarrow \mathcal{X}$ be a randomized mechanism such that for any $P$, $A(P)$ releases a sample from some $Q \in \mathcal{M}$. If $\mathcal{M}$ is an $(\alpha, \beta)$-RD Mollifer, then $A$ is an $(\alpha, \beta)$-RDP sampler.
\end{lemma}

The proof of lemma \ref{mollifier-sampler} is similar to the proof of the $\epsilon$-private sampler variant in \cite{husain2020local}. 

Since an $(\alpha, \beta)$-RD Mollifier implies an $(\alpha, \beta)$-RDP sampler, sampling from some distribution in a mollifier provides privacy. Hence, a naive privacy analysis can make use of the RD mollifier framework to derive the privacy loss, as is done in \cite{husain2020local}. For each $\overline{p}_{i}(\mathbf{x}_t), \overline{p}_{i}'(\mathbf{x}_t) \in \mathcal{M}_{(\alpha, \beta), p_{0}}$, $D_{\alpha}(\overline{p}_{i}(\mathbf{x}_t) || \overline{p}_{i}'(\mathbf{x}_t) \leq 4\beta\alpha$ due to lemma \ref{lemma:rel_conv}. Meaning, every projected distribution $\overline{p}_{i}(\mathbf{x}_t)$ in the ensemble are $(\alpha, 4\beta)$-RD samplers. Then sampling from the average of the projected distribution is still an $(\alpha, 4\beta$)-RD sampler by the Post-Processing Theorem \ref{thm:post-processing}. However, this privacy analysis is overly strict in that it's a privacy guarantee where the neighboring ensembles can differ by all models except for one, which is too strong of a privacy notion. We are interested in the opposite neighborhood condition, where two ensembles are equal for all models except for one. This allows us to take advantage of the fact that the privacy cost of sampling from a mollifier is scaled by the inverse of the size of the ensemble.

Define $p_{-i}(\mathbf{x}_t) = \frac{1}{N-1} \sum_{j\in [N]\setminus \{i\}} \lambda_j p_{j}(\mathbf{x}_t) + (1-\lambda_j) p_{0}(\mathbf{x}_t)$. To show that sampling $y_t \sim p(\mathbf{x}_t)$ is $(\alpha, \epsilon_G / T)$-RDP, i.e., $D_{\alpha}^{\leftrightarrow}(p(\mathbf{x}_t) || p_{-i}(\mathbf{x}_t)) \leq \frac{\epsilon_G}{T}$, first let's look at a special case when $\alpha=\infty$ Now 
\begin{align}
    &e^{D_{\infty}(p(\mathbf{x}_t) || p_{-i}(\mathbf{x}_t))} \nonumber \\
    &= \frac{p(y_t | \mathbf{x}_t)}{p_{-i}(y_t | \mathbf{x}_t)} \nonumber \\
    &=\frac{\frac{N-1}{N} p_{-i}(y_t | \mathbf{x}_t) + \frac{1}{N} \overline{p}_{i}(y_t | \mathbf{x}_t)}{p_{-i}(y_t | \mathbf{x}_t)} \nonumber \\
    &= \frac{N-1}{N} +  \frac{1}{N} \frac{\overline{p}_{i}(y_t | \mathbf{x}_t)}{p_{-i}(y_t | \mathbf{x}_t)} \nonumber \\
    &= \frac{N-1}{N} + \frac{1}{N} \frac{\overline{p}_{i}(y_t | \mathbf{x}_t)}{p_{0}(y_t \mathbf{x}_t)}\frac{p_{0}(y_t | \mathbf{x}_t)}{p_{-i}(y_t | \mathbf{x}_t)}\nonumber \\
    &\leq \frac{N-1}{N} + \frac{1}{N}e^{2\beta\alpha} \nonumber
\end{align} 
 for all $y_t \in V$. So if we want $D_{\infty}(p(\mathbf{x}_t) || p_{-i}(\mathbf{x}_t)) \leq \frac{\epsilon_G}{T}$ then we need set $\beta$ such that 
\begin{align}
     \frac{N-1}{N} + \frac{1}{N}e^{2\beta\alpha} &\leq e^{\frac{\epsilon_{G}}{T}} \nonumber \\
     e^{2\beta\alpha} &\leq N e^{\epsilon_G / T} + 1 - N \nonumber \\
    \beta &\leq \frac{\log\left(N e^{\epsilon_G / T} + 1 - N \right)}{2\alpha}. \label{beta_leq}
\end{align}

\noindent For the other direction: 
\begin{align}
    \frac{p_{-i}(y_t | \mathbf{x}_t)}{p(y_t | \mathbf{x}_t)} &= \frac{\frac{N}{N-1}p(y_t | \mathbf{x}_t) - \frac{1}{N-1}\overline{p}_{i}(y_t | \mathbf{x}_t)}{p(y_t | \mathbf{x}_t)}\nonumber \\
    &= \frac{N}{N-1} - \frac{1}{N-1} \frac{\overline{p}_{i}(y_t | \mathbf{x}_t)}{p(y_t | \mathbf{x}_t)}\nonumber \\
    &= \frac{N}{N-1} - \frac{1}{N-1} \frac{\overline{p}_{i}(y_t | \mathbf{x}_t)}{p_{0}(y_t | \mathbf{x}_t)}\frac{p_{0}(y_t | \mathbf{x}_t)}{p(y_t | \mathbf{x}_t)}\nonumber \\
    &= \frac{N}{N-1} - \frac{1}{N-1} e^{2\beta\alpha} \leq e^{\epsilon_G / T} \nonumber \\
    e^{2\beta\alpha} &\geq N - (N-1)e^{\epsilon_G / T} \nonumber \\
    \beta &\geq \frac{\log(N - (N-1)e^{\epsilon_G / T})}{2\alpha} \label{beta_geq}
\end{align}

for all $y_t \in V$. \noindent Equation \ref{beta_geq} is satisfied by setting $\beta$ equal to eq. \ref{beta_leq}. Thus, it suffices to find $\beta$ with order $1 < \alpha < \infty$ by working through $D_{\alpha}(p(\mathbf{x}_t) || p_{-i}(\mathbf{x}_t)) \leq \beta \alpha$, then set $\beta$ to its largest possible value so that $D_{\alpha}(p_{-i}(\mathbf{x}_t) || p(\mathbf{x}_t)) \leq \epsilon_G / T$. 

\section{Properties of Renyi Divergence and RDP}
\label{sec:useful_thms}
\begin{theorem}[Post-Processing \cite{mironov2017renyi}]\label{thm:post-processing}
   Let $A: \mathcal{D} \rightarrow \mathcal{R}$ be $(\alpha, \epsilon)$-RDP, and let $F: \mathcal{R} \rightarrow \mathcal{Z}$ be an arbitrary randomized mapping. Then $F \circ M$ is $(\alpha, \epsilon)$-RDP.
\end{theorem}

\begin{theorem}[Composition \cite{mironov2017renyi}]\label{thm:composition}
    Let $A_1, ..., A_k$ be a sequence of $(\epsilon, \alpha)$-RDP algorithms. Then the composition $A_k \circ A_{k-1} \circ ... \circ A_1$ is $(\alpha, k\epsilon)$-RDP.
\end{theorem}

\begin{theorem}[Conversion from RDP to Approximate DP \cite{balle2020hypothesis}]\label{thm:rdp-dp}
    If an algorithm $A$ is $(\alpha, \epsilon)$-RDP, then it is $(\epsilon + \log((\alpha-1)/ \alpha) - (\log\delta + \log \alpha)/(\alpha-1), \delta)$-DP for any $0 < \delta < 1$.
\end{theorem}

\begin{theorem}[Tight Privacy Amplification by Poisson Subsampling for Renyi DP \cite{steinke2022composition}]\label{thm:priacy_amp}
    Let $U \subseteq [n]$ be a random set that contains each element independently with probability $q$. For $x \in \mathcal{X}^n$ let $x_{U} \in \mathcal{X}^n$ be given by $(x_{U})_i = x_{i}$ if $i \in U$ and $(x_{U})_i = \bot$ if $i \notin U$, where $\bot \in \mathcal{X}$ is some fixed value. 
    
    Let $\epsilon: \mathbb{N}_{\geq 2} \rightarrow \mathbb{R} \cup \{\infty\}$  be a function. Let $M: \mathcal{X}^{n} \rightarrow \mathcal{Y}$ satisfy $(\alpha, \epsilon(\alpha))$-RDP for all $\alpha \in \mathbb{N}_{\geq 2}$ with resepect to addition or removal-- i.e., $x, x^{'} \in \mathcal{X}^n$ are neighboring if, for some $i \in [n]$, we have $x_i = \bot$ or $x^{'}_{i} = \bot$, and $\forall j \neq i$ $x_j = x_j^{'}$.

    Define $M^{U}: \mathcal{X}^n \rightarrow Y$ by $M^{U}(x) = M(x_{U})$. Then $M^{U}$ satisfies $(\alpha, \epsilon^{'}_{q}(\alpha))$-RDP for all $\alpha \in \mathbb{N}_{\geq 2}$ where 

    \begin{dmath}\label{eq:subsam_eps}
        \epsilon^{'}_q(\alpha) = \frac{1}{\alpha-1}\log \left((1-q)^{\alpha-1} (1 + (\alpha - 1)q) \\
        + \sum_{k=2}^{\alpha} \binom{\alpha}{k} (1-q)^{\alpha - k} q^{k} e^{(k-1) \epsilon(k)} \right). 
    \end{dmath}
\end{theorem}

\begin{theorem}[Triangle-like inequality, lemma 33.7 from \cite{steinke2022composition}]\label{thm:weak-triangle-inequality}
   Let $P, Q, R$ be distributions on $\mathcal{R}$. If $D_{\alpha}(P || Q) \leq \epsilon_1 \alpha$ and $D_{\alpha}(Q || R) \leq \epsilon_2 \alpha$ for $1 < \alpha < \infty$, then 
   \begin{equation}
       D_{\alpha}(P || R) \leq (\sqrt{\epsilon_1} + \sqrt{\epsilon_2})^{2} \alpha.
   \end{equation}
\end{theorem}

\begin{theorem}[Quasi-Convexity \cite{steinke2022composition}]\label{thm:quasi-convexity}
   Let $P, Q, P{'}, Q^{'}$ be probability distributions over $\mathcal{R}$ such that $P^{'}$ absolutely continuous with respect to $Q^{'}$. For $s \in [0, 1]$, let $(1-s)P + sP^{'}$ denote the convex combination of the distributions $P$ and $P^{'}$ with weighting $s$. For all $\alpha \in (1, \infty)$ and all $s \in [0, 1]$,
   \begin{align*}
       &D_{\alpha}((1-s) P + s P^{'} || (1-s)Q^{'} + s Q^{'}) \\ &\leq \max \{D_{\alpha}(P || Q), D_{\alpha}(P^{'} || Q^{'}) \}.
   \end{align*}
\end{theorem}

\begin{theorem}[Convexity in Second Argument \cite{van2014renyi}]\label{thm:second-convexity}
    For any order $\alpha \in [0, \infty]$ Renyi divergence is convex in its second argument. That is, for any probability distributions $P, Q, Q'$ and $s \in [0, 1]$ 
    \begin{equation*}
       D_{\alpha}(P || (1-s)Q' + sQ') \leq (1-s)D_{\alpha}(P||Q') + D_{\alpha}(P||Q).
    \end{equation*}
\end{theorem}

\section{Supplementary Experimental Figures}\label{sec:sup}
\begin{figure}[H]
    \centering
    \includegraphics[width=0.8\columnwidth]{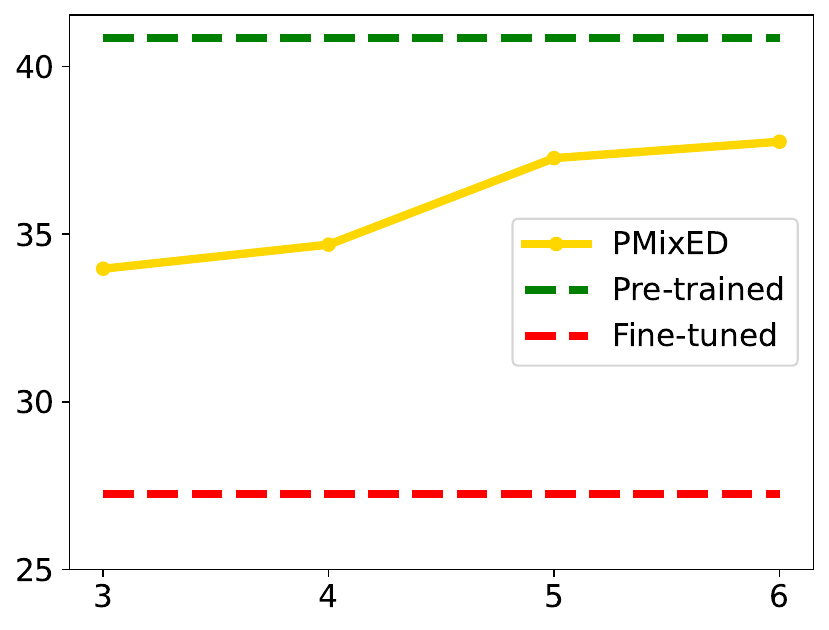}
    \caption{Alpha}
    \label{fig:alpha}
\end{figure}

Figure \ref{fig:alpha} shows the utility trade-off of $\alpha$. Hence, it is crucial to the performance of PMixED that $\alpha$ is small due to the monotonicity property of Renyi Divergence.

\section{Extended Related Works}
Another promising privacy-related notion, machine unlearning, has emerged to reduce memorization by verifiably removing learned information from a data sample without retraining a model from scratch \cite{guo2019certified, bourtoule2021machine}. Generally speaking, there are two machine unlearning notions: (1) exact unlearning, where the resulting model has completely unlearned a data sample \cite{bourtoule2021machine}, and (2) approximate unlearning, where a data point has been unlearned to some degree with high probability. It is well-known that differential privacy implies approximate machine unlearning. One recent, orthogonal work explored the use of task vectors to perform memorization unlearning of the private dataset \cite{gao2024ethos}. 

\end{document}